\documentclass[12pt,onecolumn,journal]{IEEEtran}

\usepackage{mathrsfs}
\usepackage{txfonts}
\usepackage [dvips] {graphicx}

\renewcommand\baselinestretch{2.0}

\ifCLASSINFOpdf

\else

\fi

\makeatletter
\newenvironment{tablehere}
{\def\@captype{table}} {}

 \makeatother

\begin{document}

\title{ Enhanced Compressive Wideband Frequency Spectrum Sensing for Dynamic Spectrum Access}

\author{Yipeng~Liu,
        Qun~Wan

\thanks{Yipeng Liu is with  KU Leuven, Department of Electrical Engineering (ESAT), SCD-SISTA and IBBT Future Health Department, Kasteelpark Arenberg 10, box 2446, 3001 Heverlee, Belgium;
e-mail: (yipeng.liu@kuleuven.be);.}

\thanks{Qun Wan is with the Electronic Engineering Department,
University of Electronic Science and Technology of China, Chengdu, 611731, China.
e-mail: (wanqun@uestc.edu.cn);.}

\thanks{Manuscript revised March 23th, 2012}}

\markboth{Journal Title,~Vol.~X, No.~X, Month~Year}%
{Shell \MakeLowercase{\textit{et al.}}: Bare Demo of IEEEtran.cls for Journals}

\maketitle

\begin{abstract}

Wideband spectrum sensing detects the unused spectrum holes for dynamic spectrum access (DSA).
Too high sampling rate is the main problem. Compressive sensing (CS) can reconstruct
sparse signal with much fewer randomized samples than Nyquist sampling with high
probability. Since survey shows that the monitored signal is sparse in frequency
domain, CS can deal with the sampling burden. Random samples can be obtained by the analog-to-information
converter. Signal recovery can be formulated as an L0 norm minimization and a
linear measurement fitting constraint. In DSA, the static spectrum allocation
of primary radios means the bounds between different types of primary radios are known in advance.
To incorporate this \emph{a priori} information, we divide the whole spectrum into subsections
according to the spectrum allocation policy. In the new optimization model, the minimization
of the L2 norm of each subsection is used to encourage the cluster distribution locally,
while the L0 norm of the L2 norms is minimized to give sparse distribution globally. Because
the L0/L2 optimization is not convex, an iteratively re-weighted L1/L2 optimization is
proposed to approximate it. Simulations demonstrate the proposed method outperforms others
in accuracy, denoising ability, etc.
\end{abstract}


\IEEEpeerreviewmaketitle

\section{Introduction}

Cognitive radio (CR) is a very promising technology
for wireless communication. Radio spectrum is a precious natural
resource. The fixed spectrum allocation is the major way for the
spectrum allocation now. In order to avoid interference, different
wireless services are allocated with different licensed bands.
Currently most of the available spectrum has been allocated. But the
increasing wireless services, especially the wideband ones, call
for much more spectrum access opportunities. The allocated spectrum
becomes very crowded and spectrum scarcity comes. To deal with the spectrum
scarcity problem, there are several ways, such as multiple-input
and multiple-output (MIMO) communication \cite{Alamouti_mimo}, ultra-wideband (UWB)
communication \cite{liu_uwb}, beamforming \cite{liu_bf} \cite{wang_bf}, relay \cite{wang_relay}, and so on.
Investigation demonstrates that most of the allocated bands are in very
low utility ratios \cite{fcc_report}.
CR is proposed to exploit the under-utilization of
the radio frequency (RF) spectrum. It is a paradigm in which the
cognitive transmitter changes its parameters to avoid interference
with the licensed users. This alteration of parameters is based on
the timely monitoring of the factors in the radio environment.

Spectrum sensing is one of the main functions of CR. It detects the
unused frequency bands, and then CR users can be allowed to
utilize the unused primary frequency bands. Current spectrum sensing
is performed in two steps \cite{ghasemi_spectrum_sensing}: the first
step called coarse spectrum sensing is to efficiently detect the
power spectrum density (PSD) level of primary bands; the second
step, called feature detection or multi-dimensional sensing
\cite{Yucek_spectrum_sensing_survey}, is to estimate other signal
space accessible for CR, such as direction of arrival (DOA)
estimation, spread spectrum code identification, waveform
identification, etc.

Coarse spectrum sensing requires fast and accurate power spectrum
detection over a wideband and even ultra-wideband (UWB). One
approach utilizes a bank of tunable narrowband bandpass filters. But it requires an enormous
number of RF components and bandpass filters,
which leads to high cost. Besides, the number
of the bands is fixed and the filter range is always preset. Thus
the filter bank way is not flexible. The other one is a wideband
circuit using a single RF chain followed by high-speed digital
signal processor (DSP) to flexibly search over multiple frequency
bands concurrently  \cite{sahai_spectrum_sensing}. It is flexible to
dynamic power spectrum density. High sampling rate requirement and
the resulting large number of data for processing are the major
problems \cite{tian_compressed_wideband_sensing}.

Too high sampling rate requirement brings challenge to the analog-to-digital converter (ADC). And the resulting large amount of data
requires large storage space and heavy computation burden of
DSP. Since survey shows sparsity exists in the frequency domain for
primary signal, compressive sensing (CS) can be used to
effectively decrease the sampling rate \cite{donoho_compressed_sensing} \cite{candes_robust_uncertainty_principles} \cite{candes_introduction_CS}. It
assets that a signal can be recovered with a much fewer
randomized samples than Nyquist sampling with high probability on
condition that the signal has a sparse representation.

\begin{figure}[!h]
 \centering
 \includegraphics[angle= 0, scale = 0.47]{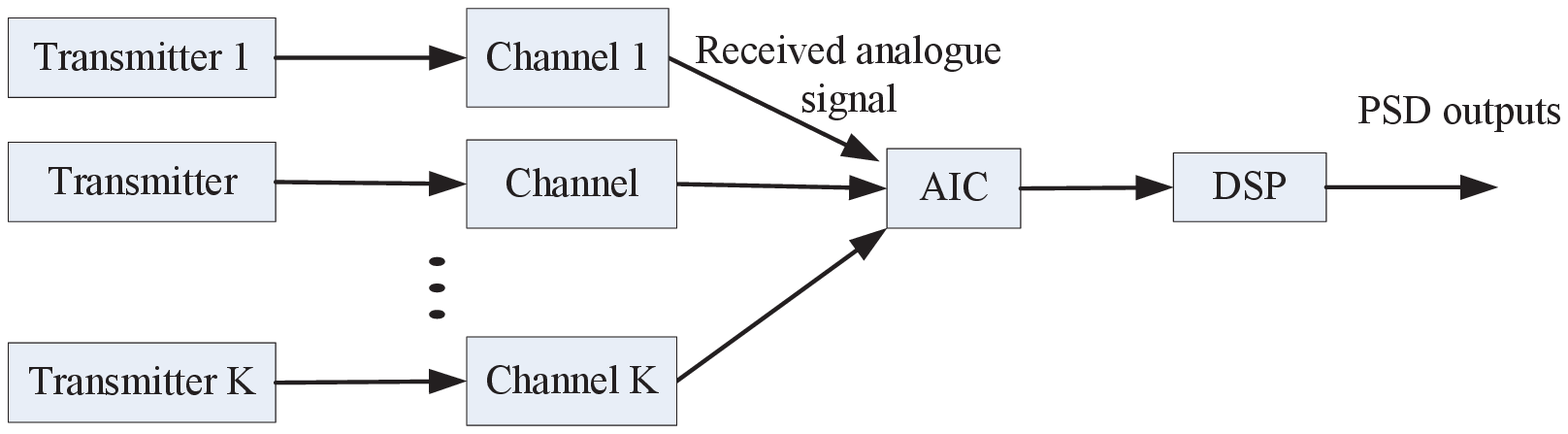}
 \caption{The proposed compressive wideband spectrum sensing structure.}
 \label{figure1}
\end{figure}

In compressive wideband spectrum sensing (CWSS), analog-to-information converter (AIC) can be taken to obtain the random
samples from analog signal in hardware as Fig. \ref{figure1} shows
\cite{laska_aic1} \cite{laska_aic2}. To get the spectrum estimation,
there are mainly two groups of methods \cite{candes_introduction_CS}. One group is convex
relaxation, such as basis pursuit (BP) \cite{chen_BP1} \cite{chen_BP2},
Dantzig Selector (DS) \cite{candes_DS} , and so on; the other is greedy algorithm, such as matching pursuit (MP) \cite{mallet_MP}, orthogonal matching pursuit (OMP) \cite{tropp_OMP}, and so on. Both
of the convex programming and greedy algorithm have advantages and
disadvantages when applied to different scenarios. A short assessment of their differences
would be that convex programming algorithm has a higher reconstruction
accuracy while greedy algorithm has less computation complexity. In
contrast to BP, basis pursuit denoising (BPDN)
has better denoising performance \cite{chen_BP2}
\cite{tibshirani_lasso}.

In this paper, the partial Fourier random samples are obtained
via AIC with the measurement matrix generated by choosing part of
separate rows randomly from the Fourier sampling matrix
\cite{laska_aic1}. Based on the random samples, a generalized
sparse constraint in the form of mixed $ \mathscr{C}_2 $/$
\mathscr{C}_1 $ norm is proposed to enhance the recovery performance
by exploiting the structure information. It encourages locally cluster
distribution and globally sparse distribution. In the constraint, the
estimated spectrum vector is divided into sections with different length
according to the \emph{a priori} information about fixed spectrum allocation. The sum of weighted
$ \mathscr{C}_2 $ norms of the sections is minimized. The weighting
factor is iteratively updated as the reciprocal of the energy in the corresponding
subband to get more democratical penalty of nonzero coefficients.
Simulation results demonstrate that the proposed generalized
sparse constraint based CWSS gets better performance
than the traditional methods in spectrum reconstruction accuracy.

In the rest of the paper, Section II gives the signal model; Section
III states the classical CWSS methods. Section IV provides the
generalized sparse constraint based CWSS methods; In section
V, the performance enhancement of the proposed method is
demonstrated by numerical experiments; Finally Section VI draws the
conclusion.

\section{Signal Model}

According to the FCC report \cite{fcc_report}, the allocated
spectrum is in a very low utilization ratio. It means the spectrum
is in sparse distribution. Recently a survey of a wide range of
spectrum utilization across 6 GHz of spectrum in some palaces of New
York City demonstrated that the maximum utilization of the allocated
spectrum is only 13.1$ \% $. It is also the reason that CR can work. Thus it is reasonable that only a small part of
the constituent signals will be simultaneously active at a given
location and a certain range of frequency band. The sparsity inherently exists in the
wideband spectrum \cite{tian_compressed_wideband_sensing}
\cite{tian_ditributed_cwss1} \cite{tian_ditributed_cwss2} \cite{elsner_cwss}
\cite{wang_distributed_cwss} \cite{polo_cwss} \cite{liu_robust_cwss}
\cite{liu_fixed_cwss}.

An $ {N \times 1} $ signal vector \textbf{x} can be
expanded in an orthogonal complete dictionary $ {\bf{\Psi }}_{N
\times N} $, with the representation as

\begin{equation}
\label{eq2_1_Signal_representation} {\bf{x}}_{N \times 1}  =
{\bf{\Psi }}_{N \times N} {\bf{b}}_{N \times 1}
\end{equation}
When most elements of the $ {N \times 1} $ vector \textbf{b} are zeros, the signal \textbf{x} is sparse. When the number of
nonzero elements of \textbf{b} is \emph{S} (\emph{S} $ \ll $ \emph{M} $
 < $ \emph{N}), the signal is said to be \emph{S}-sparse.

In traditional Nyquist sampling, the time window for sensing is $ t
\in [0,T_0 ] $. \emph{N} samples are needed to recover the frequency spectrum \textbf{r}
without aliasing, where $ T_0 $ is the Nyquist sampling duration. A
digital receiver converts the continuous signal \emph{x(t)} to a discrete
complex sequence $\textbf{y}_t$ of length \emph{M}. For illustration
convenience, we formulate the sampling model in discrete setting as it does
in  \cite{tian_compressed_wideband_sensing}
\cite{tian_ditributed_cwss1} \cite{tian_ditributed_cwss2} \cite{elsner_cwss}
\cite{wang_distributed_cwss} \cite{polo_cwss} \cite{liu_robust_cwss}
\cite{liu_fixed_cwss}:
\begin{equation}
\label{eq2_2_Signal_Sampling} {\bf{y}}_t  = {\bf{Ax}}_t
\end{equation}
where $ \textbf{x}_t $ represents an $ N \times 1 $ vector with
elements $ x_t [n] = x(t),~t = nT_0 ,~n = 1, \cdots ,N $, and
\textbf{A} is an $ M \times N $ projection matrix. For example, when
$ \textbf{A} = \textbf{F}_N $ with \emph{M} = \emph{N}, model
(\ref{eq2_2_Signal_Sampling}) amounts to frequency domain sampling,
where $ \textbf{F}_N $ is the \emph{N}-point unitary discrete Fourier
transform (DFT) matrix. Given the sample set $ \textbf{x}_t $ when $
M < N $, compressive spectrum sensing can reconstruct the spectrum
of \emph{r}(t) with the reduced amount of sampling data.

To monitor such a broad band, high sampling rate is needed. It is
often very expensive. Besides, too many sampling measurements
inevitably ask more storage devices and result in high computation burden for
digital signal processors (DSP), while spectrum sensing should be fast and accurate. CS provides an alternative to the
well-known Nyquist-Shannon sampling theory. It is a framework performing
non-adaptive measurement of the informative part of the signal
directly on condition that the signal is sparse
\cite{candes_introduction_CS}. Since it is proved that $ \textbf{x}_t $
has a sparse representation in frequency domain. We can use an $ M
\times N $ random projection matrix $ \textbf{S}_c $ to sample
signals, i.e. $ {\bf{y}}_t  = {\bf{S}}_c {\bf{x}}_t $, where $ M < N
$; $ \textbf{S}_c $ is a non-uniform subsampling or random
subsampling matrix which is generated by choosing \emph{M} separate rows
randomly from the unit matrix $ \textbf{I}_N  $.

The AIC can be used to sample the analog baseband signal \emph{x(t)}.
One possible architecture can be based on a wideband pseudorandom demodulator and a low rate
sampler \cite{laska_aic1} \cite{laska_aic2}. First we modulate the analogue signal by a
pseudo-random maximal-length PN sequence. Then a low-pass filter  follows.
Finally, the signal is sampled at sub-Nyquist rate using a traditional ADC. It can be
conceptually modeled as an ADC
operating at Nyquist rate, followed by random discrete sampling operation
\cite{laska_aic1}. Then $ \textbf{y}_t $ is obtained directly from
continuous time signal \emph{x(t)} by AIC. The details about AIC can be
found in \cite{laska_aic1} \cite{laska_aic2}. Here we
incorporate the AIC to the spectrum sensing architecture as Fig.
\ref{figure1} shows.

\section{The Classical Compressive Wideband Spectrum Sensing}

CS theory asserts that, if a signal has a sparse representation in a
certain space, one can use the random sampling to obtain the
measurements and successfully reconstruct the signal with overwhelming probability by
nonlinear algorithms, as stated in section II. The required
random samples for recovery are far fewer than Nyquist sampling.

To find the unoccupied spectrum for secondary access, the signal
in the monitored band is down-converted to baseband. The analog
baseband signal is sampled via the AIC that produces measurements at
a rate below the Nyquist rate.

Now we estimate the frequency response of \emph{x(t)} from the measurement vector
$ \textbf{y}_t $ based on the transformation equality $ {\bf{y}}_t =
{\bf{S}}_c {\bf{F}}_N^{ - 1} {\bf{r}} $, where \textbf{r} is the $ N
\times 1 $ frequency response vector (FRV) of signal \emph{x(t)}; $ \textbf{F}_N
$ is the $ N \times N $ Fourier transform matrix; $ \textbf{S}_c
$ is the $ M \times N $ matrix which is obtained by randomizing the
column indices and getting the first \emph{M} columns.

Under the sparse spectrum assumption, the FRV can be recovered by solving
the combinatorial optimization problem
\begin{equation}
\label{eq3_1_l0_recovery}
\begin{array}{c}
 {\rm{ }}{\bf{\hat r}} = \mathop {\arg \min }\limits_{\bf{r}} \left\| {\bf{r}} \right\|_0  \\
 {\rm{s}}{\rm{.~t}}{\rm{.~~}}\left( {{\bf{S}}_c^T {\bf{F}}_M^{ - 1} } \right){\bf{r}} = {\bf{y}}_t  \\
 \end{array}
\end{equation}
Since the optimization problem (\ref{eq3_1_l0_recovery}) is
nonconvex and generally impossible to solve, for its solution usually
requires an intractable combinatorial search. As it does in
\cite{tian_compressed_wideband_sensing}, BP is used to recover the
signal:

\begin{equation}
\label{eq3_2_BP-CWSS}
\begin{array}{l}
 {\rm{ }}{\bf{r}}_{BP}  = \mathop {\arg \min }\limits_{\bf{r}} \left\| {\bf{r}} \right\|_1  \\
 {\rm{s}}{\rm{.~t}}{\rm{.~~ }}\left( {{\bf{S}}_c^T {\bf{F}}_M^{ - 1} } \right){\bf{r}} = {\bf{y}}_t  \\
 \end{array}
\end{equation}
This problem is a second order cone program (SOCP) and can therefore
be solved efficiently using standard software packages.

BP finds the smallest $ \mathscr{C}_1 $ norm of coefficients among
all the decompositions that the signal is decomposed into a
linear combination of dictionary elements (columns, atoms). It is a decomposition
principle based on a true global optimization.

In practice noise exists in data. Another algorithm called
BPDN has superior denoising performance than BP
\cite{tibshirani_lasso}. It is a shrinkage and selection method for
linear regression. It minimizes the sum of the absolute values of the coefficients,
with a bound on the sum of squared errors.
To get higher accuracy, we can formulate the BPDN based compressive
wideband spectrum sensing (BPDN-CWSS) optimization model as:

\begin{equation}
\label{eq3_3_LASSO-CWSS}
\begin{array}{c}
 {\rm{ }}{\bf{r}}_{BPDN}  = \mathop {\arg \min }\limits_{\bf{r}} \left\| {\bf{r}} \right\|_1  \\
 {\rm{s}}{\rm{.~t}}{\rm{.~~ }}\left\| {\left( {{\bf{S}}_c^T {\bf{F}}_M^{ - 1} } \right){\bf{r}} - {\bf{y}}_t } \right\|_2  \le \eta_1  \\
 \end{array}
\end{equation}
where $ \eta_1 $ bounds the amount of noise in the data. The
computation of the BPDN is a quadratic programming
problem or more general convex optimization problem, and can be
done by classical numerical analysis algorithms. The solution has been well investigated \cite{tibshirani_lasso} \cite{efron_lasso}
\cite{kim_lasso1} \cite{kim_lasso2}. A number of convex optimization
software, such as cvx \cite{grant_cvx}, SeDuMi \cite{sturm_sedumi}
and Yalmip \cite{lofberg_yalmip}, can be used to solve the problem.

\section{The Proposed Compressive Wideband Spectrum Sensing}

Among the classical sparse signal recovery algorithms, BPDN achieves
the highest recovery accuracy \cite{candes_introduction_CS}.
However, it only takes advantage of sparsity. In wideband
CR application, additional \emph{a priori} information about the
spectrum structure can be obtained. The further exploitation
of structure information would give birth to recovery accuracy
enhancement \cite{liu_fixed_cwss} \cite{stojnic_block-sparse1}
\cite{stojnic_block-sparse2} . Besides, It is
well-known that the minimization of $ \mathscr{C}_0 $ norm is the
best candidate for sparse constraint. But in order to reach a convex
programming, the $ \mathscr{C}_0 $ norm is relaxed to $
\mathscr{C}_1 $ norm, which leads to the performance degeneration
\cite{candes_reweighted_l1_minimization}. Here a weighting
formulation is designed to democratically penalize the
elements. It suggests that large weights could be
used to discourage nonzero entries in the recovered FRV, while
small weights could be used to encourage nonzero entries. To get the
weighted values, a simple iterative algorithm is proposed.

\subsection{Wideband spectrum sensing for fixed spectrum allocation}

The classical algorithms reconstruct the commonly sparse signal.
However, in the coarse wideband spectrum sensing, the boundaries
between different kinds of primary users are fixed due to the static
frequency allocation of primary radios. For example, the bands 1710
- 1755 MHz and 1805 - 1850 MHz are allocated to GSM1800. Previous
CWSS algorithms did not take advantage of the information of fixed
frequency allocation boundaries. Besides, according to the practical
measurement, though the spectrum vector is sparse globally, in
some certain allocated frequency sections, they are not always
sparse. For example, in a certain time and area, the frequency
sections 1626.5 - 1646.5 MHz and 1525.0 - 1545.0 MHz allocated to
international maritime satellite are not used, but the frequency
sections allocated to GSM1800 are fully occupied. The wideband FRV
is not only sparse, but also in sparse cluster distribution
with different length of clusters. It is the generalization of the so
called block-sparsity \cite{stojnic_block-sparse1}
\cite{stojnic_block-sparse2}. This feature is extremely vivid in the
situation that most of the monitored primary signals are spread
spectrum signals.

Previous classical CWSS does not assume any additional structure on the
unknown sparse signal. However in the practical application, the signal
may have other structures. Incorporating additional structure information
would improve the recoverability potentially.

Block-sparse signal is the one whose nonzero entries are contained
within several clusters. To exploit the block structure of
ideally block-sparse signals, $\mathscr{C}_2$/$\mathscr{C}_1$
optimization was proposed. The standard block sparse constraint
(SBSC) in the form of $\mathscr{C}_2$/$\mathscr{C}_1$ optimization
can be formulated as \cite{stojnic_block-sparse1}
\cite{stojnic_block-sparse2}:

\begin{equation}
\label{eq4_1_SBSC}
\begin{array}{c}
 \mathop {\min }\limits_{\bf{r}} \left( {\sum\limits_{i = 1}^K {\left\| {{\bf{r}}_{(i - 1)d_0 :id_0 } } \right\|_2^{} } } \right) \\
 {\rm{s}}{\rm{.~t}}{\rm{.~~ }}\left( {{\bf{S}}_c^T {\bf{F}}_M^{ - 1} } \right){\bf{r}} = {\bf{y}}_t  \\
 \end{array}
\end{equation}
where \emph{K} is the number of the divided subbands; $ d_0 $ is
the length of the divided blocks. Extensive performance evaluations
and simulations have demonstrated that as $ d_0 $ grows the algorithm
significantly outperforms standard BP algorithm
\cite{stojnic_block-sparse2}.

However, in the standard $\mathscr{C}_2$/$\mathscr{C}_1$
optimization, the estimated sparse signal is divided with the same
block length, which mismatches the practical situation that the
values of the length of the spectrum subbands allocated to different radios
can not be all the same. Besides, the constraint in
(\ref{eq4_1_SBSC}) does not incorporate the denoising function.

To further enhance the performance of CWSS, the fixed spectrum
allocation information can be incorporated in the CWSS algorithm.
Based on the \emph{a priori} information about boundaries, the estimating PSD vector is divided
into sections with their edges in accordance with the boundaries of
different types of primary users by fixed spectrum allocation. In
the BPDN-CWSS, the minimization of the standard $\mathscr{C}_1$-norm
constraint on the whole FRV is replaced by the minimization of the sum
of the $\mathscr{C}_2$ norm of each divided section of the FRV to
encourage the sparse distribution globally while blocked
distribution locally. As it combines $\mathscr{C}_1$ norm and
$\mathscr{C}_2$ norm to enforce the sparse blocks with different
block lengths, the new CWSS model, in the name of
variable-length-block-sparse constraint based compressive wideband
spectrum sensing (VLBS-CWSS), can be formulated as:

\begin{equation}
\label{eq4_2_VBLS-CWSS}
\begin{array}{c}
 \mathop {\min }\limits_{\bf{r}} \left( {\left\| {{\bf{r}}_1 } \right\|_2^{}  + \left\| {{\bf{r}}_2 } \right\|_2^{}  +  \cdots \left\| {{\bf{r}}_K } \right\|_2^{} } \right) \\
 {\rm{s}}{\rm{.~t}}{\rm{.~~ }}\left\| {{\bf{y}}_t  - {\bf{S}}_c {\bf{F}}_N^{ - 1} {\bf{r}}} \right\|_2^{}  \le \eta_2  \\
 \end{array}
\end{equation}
where $ \textbf{r}_1 $, $ \textbf{r}_2 $, ... , $ \textbf{r}_K $ are
\emph{K} sub-vectors of \textbf{r} corresponding to $ d_1 $, $ d_1
$, ... , $ {d_{K - 1}} $ which are the boundaries of the divided sections. $
\eta_2 $ bounds the amount of noise in the data. It can be formulated
as:

\begin{equation}
\label{eq4_3_spectrum_division} {\bf{r}} = \left(
{\begin{array}{*{20}c}
   {\underbrace {\begin{array}{*{20}c}
   {r_1 } &  \cdots  & {r_{d_1 } }  \\
\end{array}}_{{\bf{r}}_1 }} &  \cdots  & {\underbrace {\begin{array}{*{20}c}
   {r_{d_{K - 1}  + 1} } &  \cdots  & {r_N }  \\
\end{array}}_{{\bf{r}}_K }}  \\
\end{array}} \right)^T
\end{equation}
Since the objective function in the VLBS-CWSS (\ref{eq4_2_VBLS-CWSS}) is
convex and the other constraint is an affine, it is a convex optimization problem. It can also
be solved by a host of numerical methods in polynomial time. Similar
to the solution of the BPDN-CWSS (\ref{eq3_3_LASSO-CWSS}), the optimal
\textbf{r} of the VLBS-CWSS (\ref{eq4_2_VBLS-CWSS}) can also be obtained
efficiently using some convex programming software packages. Such as
cvx \cite{grant_cvx}, SeDuMi \cite{sturm_sedumi}, and Yalmip
\cite{lofberg_yalmip}, etc.

After we get \textbf{r} from (\ref{eq4_3_spectrum_division}), power spectrum can be obtained.
Several ways can
indicate the spectrum holes, such as energy detection \cite{liu_robust_cwss}, edge
detection \cite{tian_compressed_wideband_sensing}, and so on. For example,
in energy detection we will calculate $ {\left\| {{{\bf{r}}_k}} \right\|_2} $, \emph{k} = 1, 2, ... , \emph{K}.
Comparing it with an experimental threshold, the spectrum holes for dynamic access can be clearly given.
The energy detection will be used in numerical simulations.

\subsection{Enhanced variable-length-block-sparse spectrum sensing}

In sparse constraint, $ \mathscr{C}_0 $ norm minimization is
relaxed to $ \mathscr{C}_1 $ norm at the cost of bringing the
dependence on the magnitude of the estimated vector. In the $
\mathscr{C}_1 $ norm minimization, larger entries are penalized more
heavily than smaller ones, unlike the more democratic penalization
of the $ \mathscr{C}_0 $ norm. Here in the the VLBS constraint, to
encourage sparse distribution of the spectrum in the global
perspective, the $ \mathscr{C}_1 $ norm of  a series of the $
\mathscr{C}_2 $ norm is minimized. Similarly, the dependence on the
power in each subband exits.

To deal with this imbalance, the minimization of the weighted sum of the
$ \mathscr{C}_2 $ norm of each blocks is designed to more
democratically penalize. The new weighted VLBS constraint based compressive wideband spectrum sensing
(WVLBS-CWSS) can be formulated as:

\begin{equation}
\label{eq4_4_WVBLS-CWSS}
\begin{array}{c}
 \mathop {\min }\limits_{\bf{r}} \left( {w_1 \left\| {{\bf{r}}_1 } \right\|_2^{}  + w_2 \left\| {{\bf{r}}_2 } \right\|_2^{}  +  \cdots  + w_K \left\| {{\bf{r}}_K } \right\|_2^{} } \right) \\
 {\rm{s}}{\rm{.t}}{\rm{. }}\left\| {{\bf{y}}_t  - {\bf{S}}_c {\bf{F}}_N^{ - 1} {\bf{r}}} \right\|_2^{}  \le \eta _3  \\
 \end{array}
\end{equation}
where $ \textbf{r}_1 $, $ \textbf{r}_2 $, ... , $ \textbf{r}_K $ are
defined as (\ref{eq4_3_spectrum_division}); $ \eta _3 $ bounds the
amount of noise; $ {\bf{w}} = \left[ {\begin{array}{*{20}c}
   {w_1 } & {w_2 } &  \cdots  & {w_K }  \\
\end{array}} \right]^T $.  $ w_i $ depends on $ p_i
\ge 0, {\rm{for }}~i = 1, \cdots ,K $, where $ p_i $ corresponds to
the power of the primary user exists in the \emph{i}-\emph{th} subband.

Obviously, the object function of the WVLBS-CWSS
(\ref{eq4_4_WVBLS-CWSS}) is convex. It is a convex
optimization problem. In principle this problem is solvable in
polynomial time.

To realize the WVLBS-CWSS (\ref{eq4_4_WVBLS-CWSS}), the weighting vector
\textbf{w} should be provided. As it is defined before, the
computation of the weight $ w_i $ is in fact the computation of the $ p_i $.
Here a practical way to
iteratively set the $ p_i $ is proposed. At each iteration, the $ p_i $
is the sum of the absolute value of frequency spectrum vector in the
corresponding subband. It can be formulated as:
\begin{equation}
\label{eq4_7_iteration1}
\begin{array}{c}
 p_{t,i}  = \left\| {{\bf{r}}_{t - 1,i} } \right\|_1  \\
  = \left| {r_{t - 1,~d_{i - 1}  + 1} } \right| +  \cdots  + \left| {r_{{t - 1},~d_i } } \right| \\
 \end{array}
\end{equation}
where $ {\bf{r}}_{t - 1,~i} $ is the \emph{i}-\emph{th} sub-vector as
in (\ref{eq4_3_spectrum_division}) at the (\emph{t}-1)-\emph{th}
iteration; $ r_{{t - 1},~d_{i - 1}  + 1},~ \cdots ~,~r_{t - 1,d_i }
$ are the elements of the sub-vector $ {\bf{r}}_{t - 1,i} $. After
getting the $ p_i $, the weighting vector \textbf{w} can be formulated.
Here we can get it by

\begin{equation}
\label{eq4_8_iteration2} w_i  = \frac{1}{{p_i  + \delta }}
\end{equation}
where a small parameter $ \delta  > 0  $ in
(\ref{eq4_8_iteration2}) is introduced to provide stability and to
ensure that a zero-valued component in $ p_i $ does not strictly
prohibit a nonzero estimate at the next step.

The initial condition of the recursive relation is $ w_i = 1 $, for
all $ i=1,...,K $. That means in the first step, all the blocks are
weighted equally. Along with the increase of the iteration times,
larger values of $ p_i $ are penalized
lighter in the WVLBS-CWSS (\ref{eq4_4_WVBLS-CWSS}) than smaller
values of $ p_i $. To terminate the iteration at the proper time, the
stopping rule can be formulated as

\begin{equation}
\label{eq4_9_stop_rule} \left\| {{\bf{r}}_t  - {\bf{r}}_{t - 1} }
\right\|_2  \le \varepsilon
\end{equation}
where $ {{\bf{r}}_t } $ is the estimated FRV
at the \emph{t}-\emph{th} iteration; $ \varepsilon $ bounds the
iteration residual.

The initial state of the iterative algorithm is the same with the
VLBS-CWSS (\ref{eq4_2_VBLS-CWSS}). To make a difference, The
iterative reweighted algorithm is named as enhanced variable-length-block-sparse
 constraint based compressive wideband spectrum sensing (EVLBS-CWSS).

\section{Simulation Results}

\begin{figure}[!h]
 \centering
 \includegraphics[scale = 0.47]{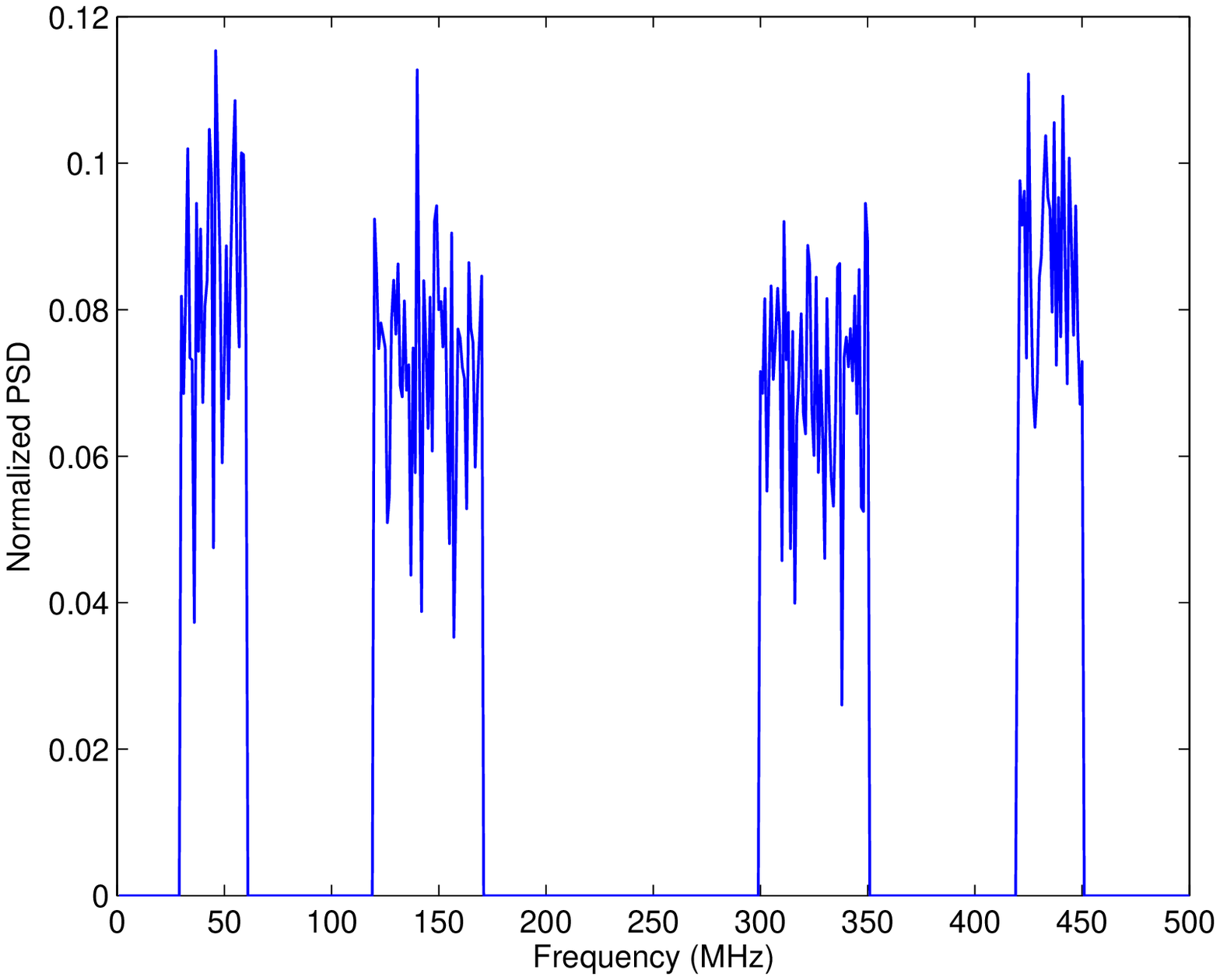}
 \caption{ The  normalized spectrum of noiseless active primary signals in the monitoring band.}
 \label{figure2}
\end{figure}

\begin{figure}[!h]
 \centering
 \includegraphics[scale = 0.47]{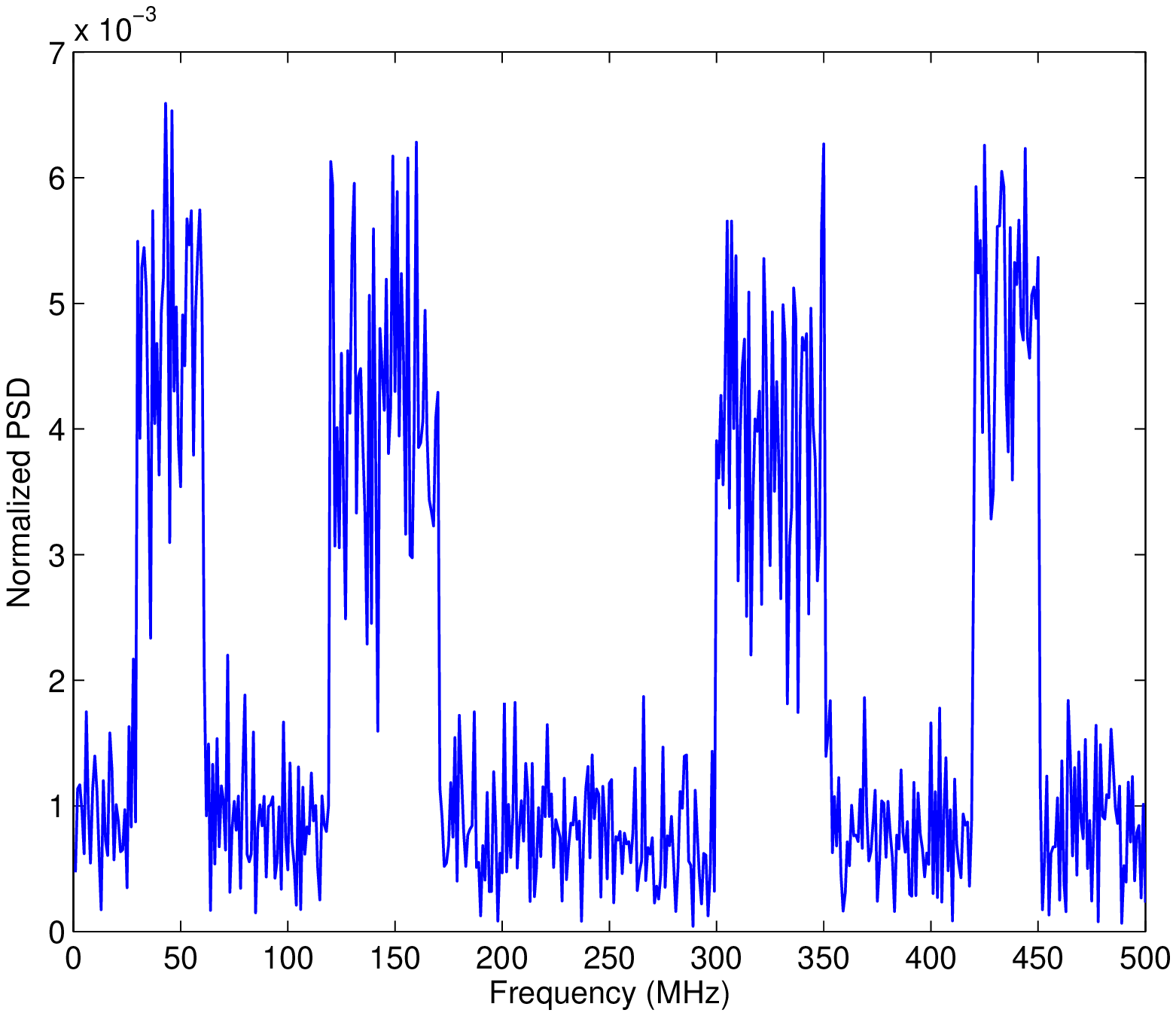}
 \caption{ The normalized spectrum of noisy active primary signals in the monitoring band.}
 \label{figure3}
\end{figure}

Numerical experiments are presented to illustrate performance improvement of the
proposed EVLBS-CWSS for CR. Here we consider a base
band signal with its frequency range from 0 Hz to 500 MHz as Fig.
\ref{figure2} shows. The primary signals with random phase are
contaminated by a zero-mean additive white Gaussian noise (AWGN)
which makes the signal to noise ratio (SNR) be 11.5 dB. Four primary
signals are located at 30 MHz - 60 MHz, 120 MHz - 170 MHz, 300 MHz -
350 MHz, 420 MHz - 450 MHz. Their corresponding frequency spectrum levels
fluctuate in the range of 0.0023 - 0.0066, 0.0016 - 0.0063, 0.0017 -
0.0063, and 0.0032 - 0.0064, as Fig. \ref{figure3} shows. Here we
take the noisy signal as the received signal \emph{x(t)}. As CS theory
suggests, we sample \emph{x(t)} randomly at the subsampling ratio 0.40 via AIC
as Fig. \ref{figure1}. The resulted sub-sample vector is denoted as
$ \textbf{y}_t $.

\begin{figure}[!h]
 \centering
 \includegraphics[scale = 0.47]{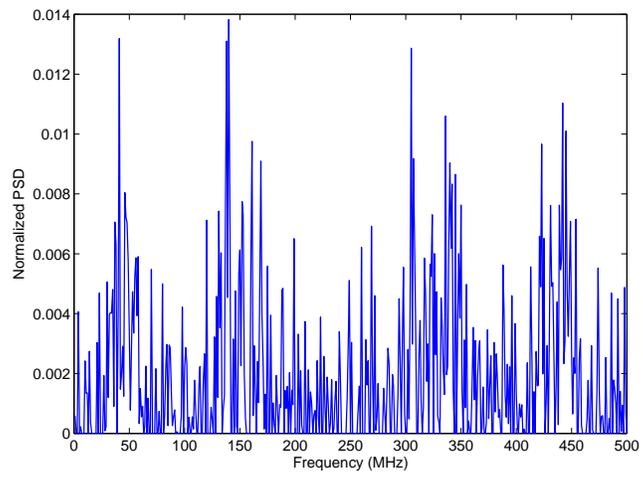}
 \caption{ The compressive wideband spectrum estimation via BPDN-CWSS.}
 \label{figure4}
\end{figure}

\begin{figure}[!h]
 \centering
 \includegraphics[scale = 0.47]{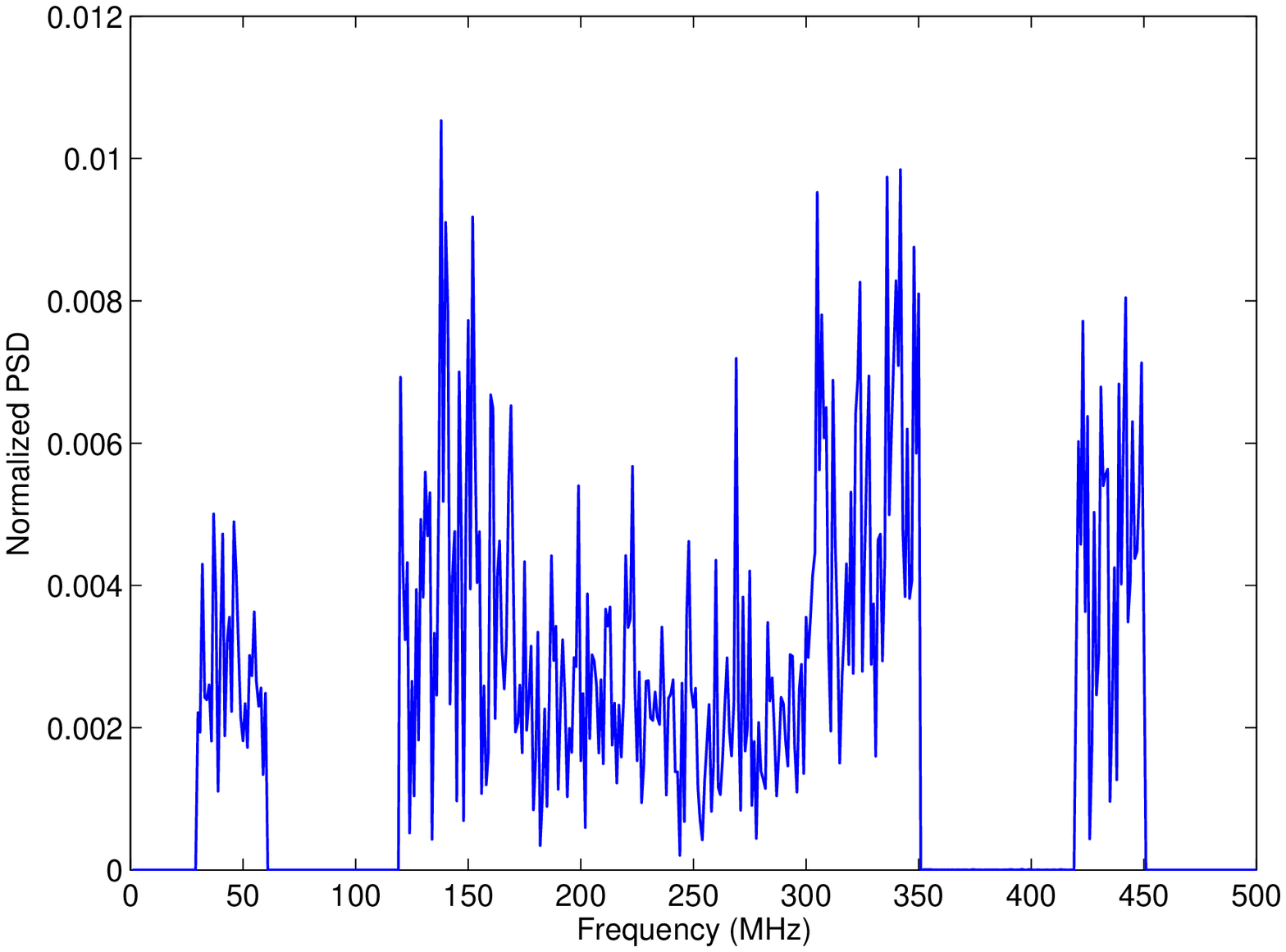}
 \caption{ The compressive wideband spectrum estimation via VLBS-CWSS.}
 \label{figure5}
\end{figure}

\begin{figure}[!h]
 \centering
 \includegraphics[scale = 0.47]{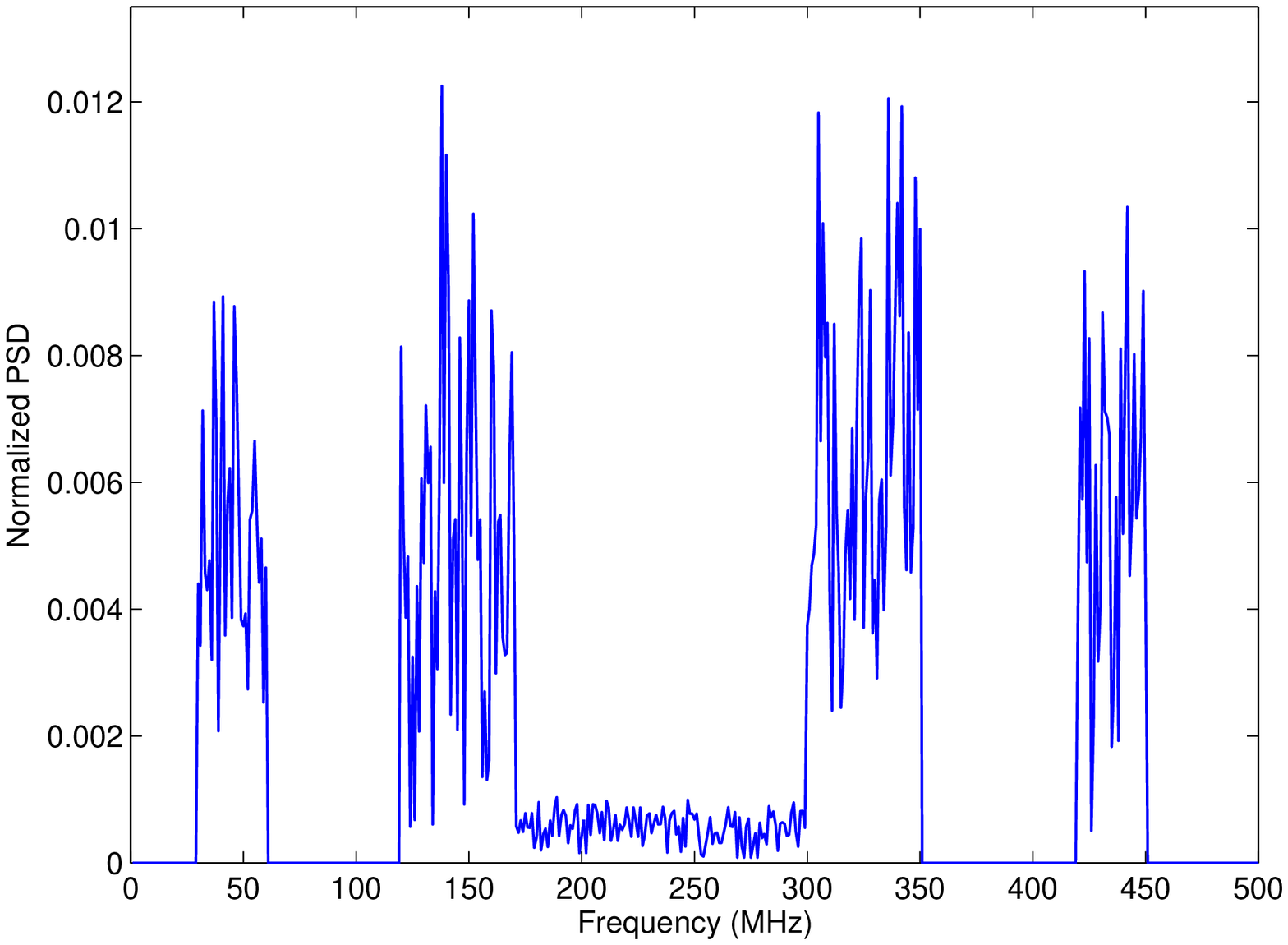}
 \caption{ The compressive wideband spectrum estimation via EVLBS-CWSS.}
 \label{figure6}
\end{figure}

To make contrast, with the same number of samples, The amplitude of frequency spectrum
estimated by different methods are given in Fig. \ref{figure4},
Fig.\ref{figure5} and Fig. \ref{figure6}. Fig. \ref{figure4} shows
the result estimated by the standard BPDN-CWSS (\ref{eq3_3_LASSO-CWSS}) where $
\eta _1 $ is chosen to be $ 0.1\left\| {{\bf{y}}_t } \right\|_2 $
with 1000 tries averaged; Fig. \ref{figure5} does it by the
VLBS-CWSS (\ref{eq4_2_VBLS-CWSS}) where $ \eta _2 $ is chosen to be
$ 0.2\left\| {{\bf{y}}_t } \right\|_2 $; Fig. \ref{figure6} does
it by the proposed EVLBS-CWSS (\ref{eq4_4_WVBLS-CWSS}) where $ \eta
_3 $ is chosen to be $ 0.2\left\| {{\bf{y}}_t } \right\|_2 $, and $ \delta $ is chosen to be 0.001.

Fig. \ref{figure6}  shows that the proposed EVLBS-CWSS gives the
best reconstruction performance. It shows that
there are too many fake spectrum points in the subbands with no
active primary signal in Fig \ref{figure4} which is given by the
standard BPDN. The noise levels of the spectrum estimated by the B-CWSS and the
VLBS-CWSS are high along the whole monitored band. For
the VLBS-CWSS, as in Fig. \ref{figure5}, it has considerable performance
improvement, but the noise level in part of the inactive subbands is
still high. Some of the estimated spectrum in the inactive subband
is a little too high. However, in Fig.
\ref{figure6}, the four occupied bands clearly show up; the noise
levels in the inactive bands are quite low; the variation of the
spectrum levels in the boundaries of estimated spectrum are quite abrupt
and correctly in accordance with the generated sparse spectrum in
Fig. \ref{figure2}, which would enhance the edge detection
performance much. Therefore, the proposed EVLBS-CWSS outperforms the
standard BPDN-CWSS and the VLBS-CWSS for wideband spectrum sensing.

Apart from the edge detection, energy detection is the most popular
spectrum sensing approach for CR. To test the CWSS performance by
energy detection, 1000 Monte Carlo simulations are done with the same
parameters above to give the results of average energy in each
section of the divided spectrum vector with the BPDN-CWSS
(\ref{eq3_3_LASSO-CWSS}), the VLBS-CWSS
(\ref{eq4_2_VBLS-CWSS}) and the EVLBS-CWSS (\ref{eq4_4_WVBLS-CWSS}).
The parameter setting is same as before. The simulated monitored
band is divided into 9 sections as Fig \ref{figure2}. The total
energy with each CWSS method is normalized. Table \ref{table1} presents
the average energy in each subband with different recovery
methods, when there are 4 active bands and the sub-sampling ratio is 0.40;
Table \ref{table2} does when there are 3 active bands and the sub-sampling
ratio is 0.40; Table \ref{table3} does when there are 3 active bands and the
sub-sampling ratio is 0.35; Table \ref{table4} when there are 2 active bands
and the sub-sampling ratio is 0.30. For the EVLBS-CWSS, it is obvious that the estimated noise
energy of inactive bands is much smaller that the other two. To
quantify the performance gain of EVLBS-CWSS against others, after
normalizing the total energy of the spectrum vectors, we define the energy
detection performance enhancement ratios (EDPER) of VLBS-CWSS and
EVLBS-CWSS against BPDN-CWSS for the \emph{k}-\emph{th} subband as:

\begin{equation}
\label{eq5_1_EDPER1} R1 \left( k \right) = \left\{
{\begin{array}{*{20}c}
   {\frac{{\left\| {{\bf{r}}_k^{VLBS} } \right\|_2^2  - \left\| {{\bf{r}}_k^{BPDN} } \right\|_2^2 }}{{\left\| {{\bf{r}}_k^{BPDN} } \right\|_2^2 }},~~{\rm{  for~active~subbands}}}  \\
   {\frac{{\left\| {{\bf{r}}_k^{BPDN} } \right\|_2^2  - \left\| {{\bf{r}}_k^{VLBS} } \right\|_2^2 }}{{\left\| {{\bf{r}}_k^{BPDN} } \right\|_2^2 }},~~{\rm{  for~inactive~subbands}}}  \\
\end{array}} \right.
\end{equation}

\begin{equation}
\label{eq5_2_EDPER2} R2 \left( k \right) = \left\{
{\begin{array}{*{20}c}
   {\frac{{\left\| {{\bf{r}}_k^{EVLBS} } \right\|_2^2  - \left\| {{\bf{r}}_k^{BPDN} } \right\|_2^2 }}{{\left\| {{\bf{r}}_k^{BPDN} } \right\|_2^2 }},~{\rm{  for~active~subbands}}}  \\
   {\frac{{\left\| {{\bf{r}}_k^{BPDN} } \right\|_2^2  - \left\| {{\bf{r}}_k^{EVLBS} } \right\|_2^2 }}{{\left\| {{\bf{r}}_k^{BPDN} } \right\|_2^2 }},~{\rm{  for~inactive~subbands}}}  \\
\end{array}} \right.
\end{equation}
where $ {{\bf{r}}_k^{EVLBS} } $, $ {{\bf{r}}_k^{VLBS} } $ and $
{{\bf{r}}_k^{BPDN} } $ represent values of estimated frequency spectrum vectors in
the \emph{k}-\emph{th} subband via EVLBS-CWSS, VLBS-CWSS and BPDN-CWSS,
respectively. These performance functions can quantify how much energy
increased to enhance the probability of correct energy detection of
the active primary bands and how much denoising performance is enhanced.
The values of EDPER in Table \ref{table1}, Table \ref{table2}, Table \ref{table3} and Table \ref{table4} clearly tell the
improvement of the proposed EVLBS-CWSS against VLBS-CWSS and
BPDN-CWSS methods.

To further evaluate the performance of EVLBS-CWSS, when the number of active bands is 4 and sub-sampling ratio is 0.40, the residuals $
\left\| {{\bf{r}}_t  - {\bf{r}}_{t - 1} } \right\|_2
 $ for 1000 Monte Carlo simulations are measured. Using the unnormalized received
signal, the measured average power of the random samples $
\textbf{y}_t  $ is 29533. From \emph{t} = 2 to \emph{t} = 8, the residuals are
361.5066, 261.6972, 55.0035, 17.9325, 15.0799, 13.4075 and 12.6189.
It shows the iteration is almost convergent at \emph{t} = 5. The
iteration would bring the increase of computation complexity, but the
performance enhancement is obvious and worthwhile.

The enhancement of spectrum
estimation accuracy qualifies the proposed EVLBS-CWSS as an
excellent candidate for CWSS.

\section{Conclusion}

In this paper, CS is used to deal with the too
high sampling rate requirement problem in the wideband spectrum
sensing for CR. The sub-Nyquist random samples is
obtained via the AIC with the partial Fourier random measurement
matrix. Based on the random samples, incorporating the \emph{a priori}
information of the fixed spectrum allocation, an improved sparse constraint
 with different block length is used to enforce
locally block distribution and globally sparse distribution of the estimated spectrum. The new
constraint matches the practical spectrum better. Furthermore,
the iterative reweighting is used to alleviate the performance
degeneration when the $ \mathscr{C}_2 $/$ \mathscr{C}_0 $ norm
minimization is relaxed to the $ \mathscr{C}_2 $/$ \mathscr{C}_1 $
one. Because the \emph{a priori} information about boundaries of different types of primary
users is added and iteration is used to enhance the VLBS
constraint performance, the proposed EVLBS-CWSS outperforms previous
CWSS methods. Numerical simulations demonstrate that the EVLBS-CWSS
has higher spectrum sensing accuracy, better denoising
performance, etc.

\section*{Acknowledgment}

This work was supported in part by the National Natural Science
Foundation of China under the grant 61172140, and '985' key projects
for excellent teaching team supporting (postgraduate) under the grant A1098522-02.

\ifCLASSOPTIONcaptionsoff
  \newpage
\fi

\begin{tablehere}
\onecolumn
\renewcommand{\arraystretch}{1.3}
\caption{The total energy in each subband with the three CWSS methods
and the values of EDPER, when there are 4 active bands and the sub-sampling ratio is 0.40. } \vskip0.2in
\begin{center}
\small \begin{tabular}{|c|c|c|c|c|c|c|c|c|c|}\hline ~ & 1 & 2 & 3 &
4& 5 & 6 & 7 & 8& 9
\\ \hline Real PSD & 0 & 0.4747 & 0 & 0.5303 & 0 & 0.5107 &
0 & 0.4823 & 0
\\ \hline BPDN-CWSS & 0.1149 & 0.3820 & 0.1752 & 0.4734 & 0.3184 & 0.4780 &
0.2333 & 0.4026 & 0.1994
\\ \hline VLBS-CWSS & 0.0000 & 0.2447 & 0.0000 & 0.5101 & 0.4220 & 0.5833 &
0.0005 & 0.4020 & 0.0000
\\ \hline EVLBS-CWSS & 0.0000 & 0.2681 & 0.0000 & 0.5396 & 0.1897 & 0.6361 &
0.0000 & 0.4431 & 0.0000
\\ \hline R1 & 100 $ \% $& -35.94 $ \% $ & 100 $ \% $ & 7.75 $ \% $ & -32.54 $ \% $ & 22.03 $ \% $ &
100 $ \% $ & 0.15 $ \% $ & 100 $ \% $
\\ \hline R2 & 100 $ \% $& -29.82 $ \% $ & 100 $ \% $ & 13.98 $ \% $ & 40.42 $ \% $ & 33.08 $ \% $ &
100 $ \% $ & 10.06 $ \% $ & 100 $ \% $
 \\\hline
  \end{tabular}
\end{center}
\label{table1}
\end{tablehere}

\begin{tablehere}
\onecolumn
\renewcommand{\arraystretch}{1.3}
\caption{The total energy in each subband with the three CWSS methods
and the values of EDPER, when there are 3 active bands and the sub-sampling ratio is 0.40.  } \vskip0.2in
\begin{center}
\small \begin{tabular}{|c|c|c|c|c|c|c|c|c|c|}\hline ~ & 1 & 2 & 3 &
4& 5 & 6 & 7 & 8& 9
\\ \hline Real PSD & 0.0000 & 0.0000 & 0.0000 & 0.5998 & 0.0000 & 0.6171 &
0.0000 & 0.5093 & 0.0000
\\ \hline BPDN-CWSS & 0.2489 & 0.1221 & 0.1704 & 0.5080 & 0.2526 & 0.5676 &
0.1637 & 0.4867 & 0.1741
\\ \hline VLBS-CWSS & 0.0000 & 0.0000 & 0.0000 & 0.5642 & 0.2544 & 0.6806 &
0.0000 & 0.3922 & 0.0000
\\ \hline EVLBS-CWSS & 0.0000 & 0.0000 & 0.0000 & 0.6029 & 0.0027 & 0.5951 &
0.0000 & 0.5313 & 0.0000
\\ \hline R1 & 100 $ \% $& 100 $ \% $ & 100 $ \% $ & 11.06 $ \% $ & -0.71 $ \% $ & 19.91 $ \% $ &
100 $ \% $ & -19.42 $ \% $ & 100 $ \% $
\\ \hline R2 & 100 $ \% $& 100 $ \% $ & 100 $ \% $ & 18.68 $ \% $ & 98.93 $ \% $ & 4.84 $ \% $ &
100 $ \% $ & 9.16 $ \% $ & 100 $ \% $
 \\\hline
  \end{tabular}
\end{center}
\label{table2}
\end{tablehere}

\begin{tablehere}
\onecolumn
\renewcommand{\arraystretch}{1.3}
\caption{The total energy in each subband with the three CWSS methods
and the values of EDPER, when there are 3 active bands and the sub-sampling ratio is 0.35.  } \vskip0.2in
\begin{center}
\small \begin{tabular}{|c|c|c|c|c|c|c|c|c|c|}\hline ~ & 1 & 2 & 3 &
4& 5 & 6 & 7 & 8& 9
\\ \hline Real PSD & 0.0000 & 0.0000 & 0.0000 & 0.5997 & 0.0000 & 0.6171 &
0.0000 & 0.5094 & 0.0000
\\ \hline BPDN-CWSS & 0.1966 & 0.1766 & 0.2002 & 0.5912 & 0.3480 & 0.4420 &
0.2573 & 0.3482 & 0.1915
\\ \hline VLBS-CWSS & 0.0000 & 0.0000 & 0.0000 & 0.8035 & 0.4190 & 0.3593 &
0.0000 & 0.2231 & 0.0000
\\ \hline EVLBS-CWSS & 0.0000 & 0.0000 & 0.0000 & 0.7572 & 0.1258 & 0.5095 &
0.0000 & 0.3887 & 0.0000
\\ \hline R1 & 100 $ \% $& 100 $ \% $ & 100 $ \% $ & 35.91 $ \% $ & -20.40 $ \% $ & -18.71 $ \% $ &
100 $ \% $ & -35.93 $ \% $ & 100 $ \% $
\\ \hline R2 & 100 $ \% $& 100 $ \% $ & 100 $ \% $ & 28.08 $ \% $ & 63.85 $ \% $ & 15.27 $ \% $ &
100 $ \% $ & 11.63 $ \% $ & 100 $ \% $
 \\\hline
  \end{tabular}
\end{center}
\label{table3}
\end{tablehere}

\begin{tablehere}
\onecolumn
\renewcommand{\arraystretch}{1.3}
\caption{The total energy in each subband with the three CWSS methods
and the values of EDPER, when there are 2 active bands and the sub-sampling ratio is 0.30.  } \vskip0.2in
\begin{center}
\small \begin{tabular}{|c|c|c|c|c|c|c|c|c|c|}\hline ~ & 1 & 2 & 3 &
4& 5 & 6 & 7 & 8& 9
\\ \hline Real PSD & 0.0000 & 0.0000 & 0.0000 & 0.7918 & 0.0000 & 0.0000 &
0.0000 & 0.6107 & 0.0000
\\ \hline BPDN-CWSS & 0.1710 & 0.0736 & 0.1733 & 0.5836 & 0.2742 & 0.1907 &
0.2341 & 0.6021 & 0.2565
\\ \hline VLBS-CWSS & 0.0000 & 0.0000 & 0.0000 & 0.7768 & 0.2151 & 0.0000 &
0.0000 & 0.5907 & 0.0000
\\ \hline EVLBS-CWSS & 0.0000 & 0.0000 & 0.0000 & 0.7697 & 0.0012 & 0.0000 &
0.0000 & 0.6387 & 0.0000
\\ \hline R1 & 100 $ \% $& 100 $ \% $ & 100 $ \% $ & 33.10 $ \% $ & 21.55 $ \% $ & 100 $ \% $ &
100 $ \% $ & -1.89 $ \% $ & 100 $ \% $
\\ \hline R2 & 100 $ \% $& 100 $ \% $ & 100 $ \% $ & 19.60 $ \% $ & 74.93 $ \% $ & 100 $ \% $ &
100 $ \% $ & 6.08 $ \% $ & 100 $ \% $
 \\\hline
  \end{tabular}
\end{center}
\label{table4}
\end{tablehere}

\end{document}